\documentclass[12pt]{article}

\usepackage{comment}
\usepackage{array}
\usepackage{booktabs}
\usepackage{mathtools}
\usepackage{graphicx}
\usepackage{xcolor}
\usepackage[version=4]{mhchem}

\usepackage{eucal}
\usepackage{scalerel}
\usepackage{subfigure}
\usepackage{parskip}
\usepackage[T1]{fontenc}

\usepackage{bm}        

\usepackage{afterpage}
\usepackage[font=scriptsize,labelfont=bf]{caption}

\usepackage{xcolor}
\usepackage{adjustbox}

\usepackage{fancyhdr}

\newcommand{\eq}[1]{\begin{equation} #1 \end{equation}}

\newcommand{\mtc}[1]{\mathrm{\mathcal{#1}}}
\newcommand{\fral}{\ \ \forall \ }
\newcommand{\trace}[1]{\mathrm{Tr} \left( #1 \right)}

\newcommand{\preprintnotice}{
\begingroup
\renewcommand\thefootnote{}
\footnotetext{
{\bf NOTE}: This manuscript is a preprint and has not undergone peer review. 

}
\endgroup
}

\usepackage{graphicx}
\usepackage[space]{grffile}
\usepackage{latexsym}
\usepackage{textcomp}
\usepackage{longtable}
\usepackage{tabulary}
\usepackage{booktabs,array,multirow}
\usepackage{amsfonts,amsmath,amssymb}

\usepackage[numbers, compress]{natbib}

\usepackage{url}
\usepackage{hyperref}
\hypersetup{colorlinks=false,pdfborder={0 0 0}}
\usepackage{etoolbox}
\makeatother
\newif\iflatexml\latexmlfalse

\AtBeginDocument{\DeclareGraphicsExtensions{.pdf,.PDF,.eps,.EPS,.png,.PNG,.tif,.TIF,.jpg,.JPG,.jpeg,.JPEG}}

\usepackage[utf8]{inputenc}
\usepackage[english]{babel}
\usepackage{siunitx}

\iflatexml

\else

\usepackage{authblk}

\begin{document}

\fi



\title{Shannon and R\'enyi entropies of molecular densities: insights into extensivity and the incomplete description of electron correlation}
\date{March 5, 2026}

\author[1]{Diogo J. L. Rodrigues\thanks{Email: diogojlrodrigues@gmail.com}}
\author[1]{Evelio Francisco}
\author[1]{Ángel Martín Pendás}

\affil[1]{Department of Physical and Analytical Chemistry, Faculty of Chemistry, University of Oviedo, Av. Julian Claveria 8, 33006 Oviedo, Spain}


\maketitle
\selectlanguage{english}
\preprintnotice 

\begin{abstract}
{\small 
In this work, we investigate the reliability of information-theoretic measures based on the electron-density and shape-function, specifically Shannon and R\'enyi entropies, as descriptors of electronic correlation. By establishing a rigorous decomposition of these entropic measures into additive and nonadditive contributions, supported on a Mulliken-like atomic partition of molecules, we systematically analyze the asymptotic behavior of the entropies at the infinite-internuclear-distance limit to assess the problem of static correlation and extensivity. 
Our algebraic and numerical analysis reveals several flaws in the use of these density-based descriptors. We demonstrate that for minimal-basis and different theoretical levels, the Shannon and R\'enyi entropies fail to encode the amount of static correlation conveyed by the underlying wavefunction. 
Conversely, shape-function Shannon entropies and R\'enyi entropies (for $\alpha \neq 1$)
violate extensivity. 
In larger basis sets, uncorrelated Hartree-Fock densities consistently overestimate entropy compared to sufficiently correlated (e.g., full-valence-CAS) densities. Moreover, the entropies for insufficiently correlated methods violate extensivity. These findings indicate that electron-density-based measures are insufficient for capturing static correlation, suggesting that robust entropic descriptors should be constructed from higher-dimensional Hilbert-space objects.

}

\end{abstract}%

\section{Introduction}

The foundations of modern information theory trace back to Claude Shannon’s seminal 1948 paper, “A Mathematical Theory of Communication” \cite{SHANNON}, which laid the groundwork for broad applications across the sciences \cite{brillouin2013science}. In physics, information theory  has played a unifying role by enabling the derivation of fundamental laws via the Extreme Physical Information principle \cite{frieden2007physics}, a framework that has, for instance,  provided a  fresh view of statistical thermodynamics based in Jaynes’ Maximum Entropy principle \cite{Jaynes1957620}, which can justify Boltzmann's distribution  with the thermodynamic entropy reducing to the Shannon entropy of our information about the system. Information theory has also found immense applications in other fields, such as molecular biology \cite{gatenby2007information,segal2003use}.

The increasing widespread use of informational measures in theoretical chemistry can be traced to their offering of unique insights into the fundamental aspects of molecular systems and chemical processes. 
Informational entropic measures have recently been used as tools to provide different frameworks to understand electronic structure \cite{SHANNON1,SHANNON2,SHANNON3,Ho19945149,Tripathi19924385,Guevara20030125071,Guevara20037030,Ho199415,Nagy1996323, Noorizadeh20104742, Ayers2006370, Sagar2005} 
and molecular reactivity \cite{nalewajskibook,welearegay2014information, MolinaEspritu2015}, to probe electron correlation \cite{AlivertiPiuri2024,stein2017measuring,Boguslawski2016,Sagar20029213} and molecular bonding \cite{barrales2024shannon,tubman2012renyi,nalewajski2000entropic}, and to build atomic, molecular complexity \cite{angulo2008atomic, bottcher2016additive, nagy2011renyi}, 
and molecular similarity measures \cite{Ho19985469,lin1996correlation}. Fischer information has also recently seen widespread interest by the chemistry community \cite{Liu2014, nalewajskibook, Ludea2019, nalewajski2010additive, nalewajski2010use, esquivel2011}, and the \textit{information conservation principle} has also been raising interest \cite{Liu2014}.

Information-theoretic tools based on the electron density are also central to the development of theoretical  chemistry, especially regarding the definition of "quantum atoms" within molecules. For instance, the Hirshfeld partitioning method is rooted in an entropy-deficiency functional that reflects informational deviations between promolecular and actual electron densities \cite{nalewajski2000information, Ayers2006370}. This framework has led to the formulation of continuity equations for position-space, momentum-space, and phase-space descriptors of molecular equilibrium \cite{nalewajski2016, nalewajski2015, nalewajski2018}.  The so-called "information-theoretic atoms" are remarkably robust: their properties remain consistent regardless of the chosen information metric, and they exhibit thermodynamic-like behavior \cite{
nalewajski2001information}. In chemical theory, these overlapping atomic domains, defined relative to the free atoms of a promolecule and embedded in the molecular environment, serve as natural building blocks of molecules. 
The Hirshfeld partition has also proven useful in visualizing and analyzing electron distribution among atomic fragments. Its applications span charge distribution analysis \cite{Harrison2009,Mandado2004}, crystallography \cite{Buinsk2019}, and the study of non-covalent interactions \cite{Gaur2024,Zhang2011} 
which are essential for understanding molecular stability and intermolecular recognition.

In the context of entropic informational measures in theoretical chemistry, we find two major classes. One is based on quantum information theory, mostly built around the von Neumann entropy ansatz. Examples are entropies that use an atoms in molecules approach \cite{VanHende2024},  maximally-entangled atomic orbitals \cite{ding2025entanglement}, or  quantum-entropic measures of orbital entanglement \cite{Boguslawski2014,schilling2021orbital,stein2017measuring, Boguslawski2016}.
The second is based on classical information theory and the  use of Shannon entropies based on the electron density. One example is the use of entropic approaches in conceptual density functional theory \cite{Geerlings2020}. Extensions of classical information theory to the pair density have been recently proposed \cite{Zhao2025-2}, as well as the introduction of the concept of \textit{Information Energy} in molecular studies \cite{He2024,He2025}. We will focus on this second class in this work, noticing that reviews on both approaches can be found, for instance,  in \cite{Zhao2025}  and  \cite{AlivertiPiuri2024}.

The Shannon entropy ($S_\rho$) \cite{SHANNON} is a cornerstone of information theory, and to no surprise, is also widely applied in theoretical chemistry \cite{nalewajski2014entropic}, among a wide range of  fields \cite{levine1986theory}. It measures the spread of probabilities of an appropriately defined distribution  function and reveals the \textit{uncertainty} of the outcome described by said distribution.  Since in the Born interpretation of quantum mechanics the electron density, $\rho$,  can be understood as a probabilistic distribution, its Shannon entropy  is accordingly  defined as:
\eq{
S_{\rho} = -\int \rho({\bf r}) \log \rho({\bf r}) d{\bf r}.
}
In contrast with the also widely used Jaynes entropy  obtained from the natural orbital occupation numbers or the Von Neumann entropy of the full molecular density matrix, the electron-density Shannon entropy helps measure the spread, shape, and symmetry of $\rho$. The base of the logarithm is typically taken as $2$, although base-$10$ cases are also occasionally found.  Its units are typically called \emph{nats} or \textit{bits}, respectively. 
To give this entropy a higher resemblance to the usual normalized probability distributions that occur in information theory, one often writes it in terms of the shape function as defined by Parr \textit{et al.}, $\sigma({\bf r}) = \rho({\bf r})/N$ \cite{Parr2005}, $N$ being the number of electrons, as:
\eq{
S_{\sigma} = -\int \sigma({\bf r}) \log \sigma({\bf r}) d{\bf r}.
}

The Shannon entropy has been widely employed to characterize the electronic structure of both atomic and molecular systems. For atomic systems, it has been applied to unravel the shell structure, to quantify basis set effects, or to gauge electron correlation effects in single atoms or ions \cite{SHANNON1, SHANNON2, SHANNON3, Ho19945149, Tripathi19924385, Guevara20030125071, Guevara20037030, Noorizadeh20104742, Sagar2005}. In simple diatomic systems, such as two-electron atoms and molecules, it has been used to examine how entropic measures change with bond distance and electron correlation \cite{lin2015shannon} and, for general molecular systems, it has been employed to follow trends in electronic delocalization and reactivity \cite{Ho199415, Nagy1996323, barrales2024shannon, FloresGallegos2019, FloresGallegos2023, Matrodi2020}. 
Many other applications include the assessment of basis set quality \cite{Ho19945149, Ho199415, Gadre1985970, Gadre19852602}, providing a sensitive diagnostic for basis set completeness. It has been used to monitor features related to electron correlation and entanglement \cite{Wang2025,Guevara2005, Hò199810620}, to track changes along reaction paths \cite{Ho2000376}, and to explore fundamental concepts like aromaticity \cite{Noorizadeh20104742}. In this case, the so-called Shannon aromaticity index, for instance, is capable of capturing aromaticity trends solely from the electron density and a quantum theory of atoms in molecules (QTAIM) partitioning \cite{bader}. In this regard,  He and coworkers \cite{he2020towards} reported strong correlations between the Shannon entropy in transition-metal systems and various electronic properties associated with  all-metal aromaticity

In a set of   detailed case studies, Hô et al. \cite{Hò199810620} analyzed the Shannon entropy for the water molecule across a range of geometries and levels of theory. They observed that better correlated electronic structured methods yield higher entropies, and that position- and momentum-space entropies exhibit opposite trends. Notably, dynamic correlation was found to increase the real-space entropy, while static correlation contributed more significantly in momentum space. In atoms, Guevara and coworkers \cite{Guevara2005} found that the local correlation entropy closely resembles the traditional correlation function and is dominated by contributions from valence electrons, although core electrons also play a role. 

Since in Density Functional Theory (DFT) the electron-density should unequivocally define all molecular properties, many forms of information-theoretic (IT) techniques have been applied to DFT densities \cite{rong2020information, liu2016information}. Relations between generalized (information-)entropic measures  and DFT local functionals have been pointed out \cite{flores2017generalized,flores2022possible}, as well as relations between the various different IT measures, such as between different entropies and Fischer information \cite{rong2020information}.

The real space electron-density Shannon entropy $S$ has also been linked to the electron correlation energy of a weakly inhomogenous electron gas in atoms under the scope of the Thomas-Fermi model as \cite{grassi2011relationship}:
\eq{
E_{corr} = A S_{\rho} + BN
}
where $A$ and $B$ are constants and $N$ is the electron number. This points towards a possible promising general relation between the entropy and the correlation energy for many-electron systems. While this idea is interesting and makes sense under the assumption of a clear relation between localization and correlation, we will see in this work that such an assumption may fail.  Some features of chemical bonding (deeply intertwined with correlation \cite{torre2003bond,casals2019bond}) cannot be captured by standard electron density-based entropic informational measures.
Furthermore, the Collins conjecture \cite{collins1993entropy}, formulated in 1993, relates the Jaynes entropy ($S_{Jaynes}$) of the one-electron reduced density matrix  (1RDM) to the correlation energy of many-electron systems, as:
\begin{equation} \label{eqI2}
S_{Jaynes}= -\sum_i n_i \log n_i = k E_{corr}
\end{equation}
where $k$ is a constant and the $n$'s are the natural occupations of  the  $\rho(\bf{r},\bf{r}')$ 1RDM . This conjecture has been both supported and contradicted on different conditions \cite{Ramrez1997,Hò199810620},  being later  extended by Paul Ziesche \cite{ziesche1995correlation} for the momentum density of the uniform electron gas.
Recent research has proposed a modified Collins conjecture which would relate (up to a constant) the non-freeness entropy \cite{gottlieb2007properties} of the electronic system with its intrinsic correlation energy \cite{cioslowski2024constraints}, though arguments against this modified conjecture have also been given \cite{cioslowski2024constraints}. We also notice that although the Jaynes and the electron-density Shannon entropy are different, there exist relations  between them as described by Sagar \cite{Sagar20029213} and  Ramirez \textit{et al} \cite{ramirez1998amount}.  Simply speaking,  Jaynes' entropy is the Von Neumann entropy of the 1-RDM, while the entropy of the electron density is just that of its diagonal part when  expressed in the position basis. Unless the 1-RDM is diagonal, they will differ, and the use of the electron density will ignore the effects of delocalization, entanglement, phase information, coherence, and momentum-space spread, although some of the missing momentum-space information can be restored by using  the position and momentum space entropies simultaneously \cite{Gadre1985, Gong2012}.
In 2015, Delle Site gave numerical results \cite{DelleSite20151396} which support a possible extension of the Collins conjecture from Jaynes' entropy to the Shannon entropy of the electron density. The author argues that it may be the case that the functional dependence of the correlation energy on the electron-density may be its Shannon entropy, at least to a first order approximation.

The Rényi entropy \cite{renyi}, see below,  is a generalization of Shannon entropy that loosens the fourth Kinchin axiom of strong additivity \cite{kinchin}. It has had noteworthy applications in the informational study of atomic structure \cite{Romera2008,Nagy2008,Nasser2017, LIU2015}, molecular bonding \cite{tubman2012renyi} ,   electron correlation \cite{FloresGallegos2021}, reactivity \cite{LIU2015}, as well as in evaluating atomic and molecular complexity \cite{nagy2011renyi, SnchezMoreno2014, Antoln2009, bottcher2016additive}.
Generalized informational entropies, and in particular Rényi entropies, can be used to define quantum uncertainty relations \cite{BialynickiBirula2011, BialynickiBirula2006}.

Of utmost importance to chemistry is the fact that, while the Shannon entropy is additive on \textit{independent} probabilities, it is generally not on the sum of additive terms, such as those that arise as additive contributions to the probabilities. Such terms are relevant when one works with the entropy of the electron density. For example, when trying to resolve electronic-density entropies into their atomic or orbital contributions, one does not obtain general direct transformations into sums of atom- or orbital-resolved terms. This is an important factor in informational molecular electronic structure theories \cite{nalewajski2014entropic,nalewajski2010additive,nalewajski2000entropic, nalewajski2013} that is closely related to the nature of interacting atoms in a molecule \cite{nalewajski2010use}. Similar situations arise when using generalized entropic measures such as those of Rényi.

Provided that, as shown, information-theoretic entropies based on the electron density  have found their way into chemistry, we show in this work, by means of very simple examples, that their practical power to uncover the effects of electron correlation and to reveal chemical bonding effects, arguably the cornerstone of chemistry, may have been overrated. To that end, we study in detail minimal and extended basis-set models of two-electron homodiatomics at the single-determinant Hartree-Fock (HF), Heitler-London (HL), and multideterminant (Full Configuration Interaction (FCI) for minimal-basis, Complete Active Space Self-Consistent Field (CAS) for extended basis) levels, paying special attention to their dissociation processes, where the different levels of theory  differ in the amount of static and dynamic electron correlation actually introduced.  To understand their different behaviors, we derive a partition, both for the Shannon and Rényi  entropies, into atomic-additive and nonadditive contributions. We also touch upon the atomic-like partition of entropies in the general case, obtaining infinite distance limits valid for both electron-density and shape function entropies.

Infinite distance limits are also  devised for the Rényi entropies of any order $\alpha$, for which a persisting nonadditive contribution is found additionally to a non-extensive form of the atomic limit.
Our algebraic results, also supported numerically throughout the manuscript, show that information-theoretic entropic measures based on practical, approximate electron densities cannot capture either chemical bonding or static electron correlation effects. 
Finally, a few conclusions follow.

\section{Shannon entropy}

\subsection{Partition of the Shannon entropy into atom-center contributions} \label{sec2.1}

Partitions of molecules into atoms, an active field of theoretical chemistry, have been extensively researched \cite{popelier2000atoms, MartnPends2023, HeidarZadeh2017}. For the purposes of this work, we  use a Mulliken-like decomposition of  the spatial electron-density into atomic/interatomic additive contributions obtained from an atom-centered primitive basis-set expansion with  $\mathcal{L}$ terms
\eq{ \label{eqx1}
\rho ({\bf r}) = \sum_{i=1}^{\mtc{L}} \sum_{j=1}^{\mtc{L}} c_{ij} \phi_i ({\bf r})  \phi_j ({\bf r}), 
}
where $i$ and $j$ are function indices. We consider $\mtc{L_A}$ functions for each center $\{ \mtc{A} \} = A,B,\dots \equiv A,A_1,A_2 \dots A_{\mtc{N}}$, and rewrite the above expression into sums over the different $\mtc{N}$ atomic center pairs $\mtc{AB}$ (\textit{a là} Mulliken \cite{Mulliken}):
\eq{
\rho ({\bf r})  =
\sum_{\mtc{A}=A}^{A_\mtc{N}} \sum_{\mtc{B}=A}^{A_\mtc{N}} \rho^{\mtc{A B}} ({\bf r}), \label{eqx2}
}
with $\rho^{\mtc{A B}} ({\bf r}) \equiv  \sum\limits_{i\in \mtc{A}} \sum\limits_{j\in \mtc{B}} c_{ij} \phi_i^{\mtc{A}} ({\bf r})  \phi_j^{\mtc{B}} ({\bf r})$ - where $\sum\limits_{i \in \mtc{A}}$ is the sum restricted to the orbitals that are centered on $\mtc{A}$ -
 are the center-resolved contributions to the electron-density, summing all function pairs that match the designated atom pair $\mtc{AB}$.  Although Mulliken's partitioning is known to display severe deficiencies when extended basis sets are used, its behavior in the case of minimal basis sets matches chemical expectations and allows for  simple algebraic manipulations that suit our needs. 

Although we work directly with the electron-density here, everything is unchanged if one divides it by the number of electrons $N$, obtaining the shape function $\sigma ({\bf r}) = \rho ({\bf r})/N$. We can thus use $\rho({\bf r})$ as a placeholder for the electron-density  or the shape-function. The latter entails more similarity with classical entropies of probability distributions due to normalization, also allowing for a better comparison in the case of molecules with different number of electrons. Notice that both entropies are related by $S_{\rho}= - \int \rho({\bf r}) \log \rho ({\bf r}) d{\bf r} = - N \int \sigma({\bf r}) \log \sigma ({\bf r}) d {\bf r} \ - \ N\log{N} = N{S_{\sigma}} \ - \ N\log{N}$.

Shape-function normalization also ensures it obeys 
$\sigma({\bf r}) \leq 1 \fral {\bf r}$, which is a necessary constraint to avoid negative contributions to the entropy. 
Although it is unlikely that $S_\rho$ does not obey non-negativity for general systems, %
 it is not guaranteed by mathematical construction, and while
this generally does not pose a problem in an entropic analysis of non-normalized electron-densities, $S_\sigma$ still appears to be a cleaner mathematical object than $S_\rho$. However, as we shall see, using the shape function has some other significant drawbacks, especially as extensity is considered  in the study of the dissociation limit.   

Using this partition as a starting point, the Shannon entropy density $- \rho({\bf r}) \log \rho(\bf r)$ of the electron-density cannot be directly partitioned into additive atom-pair terms as in Equation (\ref{eqx2}), due to the non-additive nature of the logarithm under additive terms in its argument. As such, we subtly rewrite the logarithm of the electron-density into its additive contributions and remaining, nonadditive remainders:  
\eq{ \label{eqx.1}
- \log \rho({\bf r}) = - \log \Bigg( \sum_{\mtc{A}=A}^{A_\mtc{N}} \sum_{\mtc{B}=A}^{A_\mtc{N}} \rho^{\mtc{A B}} ({\bf r}) \Bigg) =  - \sum_{\mtc{A}=A}^{A_\mtc{N}} \sum_{\mtc{B}=A}^{A_\mtc{N}} 
\frac{\rho^{\mtc{A B}} ({\bf r})}{\rho({\bf r})}  \log \rho^{\mtc{A B}} ({\bf r}) + \sum_{\mtc{A}=A}^{A_\mtc{N}} \sum_{\mtc{B}=A}^{A_\mtc{N}} \frac{\rho^{\mtc{A B}} ({\bf r})}{\rho({\bf r})} \log \frac{\rho^{\mtc{A B}} ({\bf r})}{\rho({\bf r})} 
}
This is an expansion akin to writing the thermodynamic free-energy (the logarithm of the partition function) in terms of average energy and entropy contributions. 
Since negative or imaginary contributions may emerge in some special cases, one should change this definition into one that uses only the logarithm of absolute values, which we will define as $\log'(a) \equiv \log|a|$. This does not alter the validity of the rewritten expression, as one can easily check. %
Multiplying by the electron-density outside the logarithm, we obtain an atom-resolved entropy density  in terms of its center additive ($S^{add}_{\rho}({\bf r})$) and nonadditive ($S^{nadd}_{\rho}({\bf r})$) contributions:
\begin{align}
S_{\rho}({\bf r}) \equiv - \rho({\bf r}) \log \rho({\bf r}) &=& \underbrace{ - \sum_{\mtc{A}=A}^{A_\mtc{N}} \sum_{\mtc{B}=A}^{A_\mtc{N}} \rho^{\mtc{A B}} ({\bf r}) \log' \rho^{\mtc{A B}} ({\bf r}) } \ &+ \ \underbrace{\sum_{\mtc{A}=A}^{A_\mtc{N}} \sum_{\mtc{B}=A}^{A_\mtc{N}} \rho^{\mtc{A B}} ({\bf r}) \log' \frac{\rho^{\mtc{A B}} ({\bf r})}{\rho({\bf r})} } \notag \\
&\equiv& S^{add}_{\rho}({\bf r}) \ \ \ \ \ \ \ \ \ \ \ \ &- \ \ \ \ \ \ \ \ \ \ \ \ \ \ S^{nadd}_{\rho}({\bf r})
\end{align}
The nonadditive part of the entropy measures the coupling (deviation from additiveness) between the different same-center ($\mtc{A}=\mtc{B}$) and center-overlap ($\mtc{A} \neq \mtc{B}$) contributions. $S^{nadd}_{\rho}({\bf r})/\rho({\bf r})$ can be identified as the (discrete) Shannon entropy of the sum $\sum\limits_{\mtc{A},\mtc{B}} \rho^{\mtc{AB}} ({\bf r})$.
We define the nonadditive entropy such that it is always positive (as expected for an entropy), which, since it is subtracted from $S^{add}_{\rho}({\bf r})$, is always a term that lowers the entropy when compared to its purely additive component. In other words, it corrects for the overestimation of the additive part of the entropy as compared to the actual molecular entropy.


\subsection{Case study - Partition of the entropy of two-electron homonuclear diatomic molecules at different levels of theory with minimal-basis sets}

In this section, we illustrate the partition of the entropy derived to the case of a minimal basis set two-electron homodiatomic at the HF, HL, and FCI levels. This textbook example still allows for  qualitative conclusions that remain valid at higher levels of theory. It also serves as an example to build insight into the partition proposed, and a toy model to extract infinite distance limits. As we will see in the next section, these limits are general for any molecule, and to some extent independent of the minimal-basis approximation.

\subsubsection{The  Hartree-Fock single-determinant} \label{sec2.2.2}

For a closed-shell two-electron system built from one atomic orbital per center, we have a symmetry-fixed single set of bonding and antibonding molecular orbitals (see SI section 1):
\eq{
\Psi_{+/-} ({\bf r}) = \frac{\phi^A ({\bf r}) \pm \phi^B ({\bf r}) }{\sqrt{2(1 \pm S)}} 
}
These are the traditional $\sigma_g$ (+) and $\sigma_u$ (-) functions.
The ground-state HF single determinant based on these molecular orbitals is:
\eq{
\Psi({\bf x_1},{\bf x_2}) = \sigma_g({\bf r_1})  \sigma_g({\bf r_2}) \times \frac{1}{\sqrt{2}} \Big( \alpha(1) \beta(2) - \beta(1) \alpha(2) \Big)}
where $\bf{x_i}$ are spin-position coordinates. 
We now apply the partition developed in the previous section to the density of the molecule described by this determinant.
One can easily identify the center-pair densities (see SI section 2)
\begin{align}
&\rho^{\mtc{AA}}({\bf r}) =  \frac{\phi^{\mtc{A}} ({\bf r}) ^2}{1 + S} , \quad \mtc{A}=A\ \mathrm{or} \ B \label{eqz1}\\
&\rho^{AB}({\bf r}) = \rho^{BA} =
\frac{ \phi^A ({\bf r}) \phi^B ({\bf r}) }{1 + S} \label{eqz2},
\end{align}
with corresponding entropy density: %
\eq{ \label{eqz3}
S_{\rho} ({\bf r}) =  \underbrace { -\frac{\phi^A ({\bf r})^2}{1+S} \log \frac{\phi^A ({\bf r})^2}{1+S} }_{S^{net,B}_{\rho}} \ \ \ \underbrace{ -\frac{\phi^B ({\bf r})^2}{1+S} \log \frac{\phi^B ({\bf r})^2}{1+S} }_{S^{net,B}_{\rho}} \ \ \  \underbrace{ -2\frac{\phi^A({\bf r}) \phi^B ({\bf r})}{1+S} \log \Big\{ 2\frac{\phi^A({\bf r}) \phi^B ({\bf r})}{1+S}  \Big\} }_{S_{\rho}^{overl,AB}}\ - \ S^{nadd}_{\rho}({\bf r}). 
}
The additive part has been decomposed, as can generally be done, into the sum (for all atoms) of atomic net $S^{net,\mtc{A}}_{\rho}$ and overlap (for all pairs of atoms) $S_{\rho}^{overl,\mtc{AB}}$ contributions. The nonadditive entropy is equal to the additive part explicitly written above, except we divide by $\rho({\bf r})$ in the argument of the log.
Thus, if the $\phi^{A/B}$ orbitals are taken to be $s$ orbitals, then the molecular entropy differs from the free atom entropies solely by the overlap and nonadditive terms (and the overlap $S=\langle \phi^A\vert\phi^B \rangle$ in the denominator of the atomic net contributions). If some other reconstruction of the 1$s$ orbital is used, we have an extra difference in entropy between the atom-in-the-molecule and the free atom.

We now examine the asymptotic behavior in the limit of infinite interatomic separation, $R = |\mathbf{R}| \equiv |\mathbf{R}_A - \mathbf{R}_B| \to \infty$. In this regime, 
$S$ tends to zero, and products of orbitals centered on different nuclei also vanish pointwise: $S \to 0$ and $\phi^A(\mathbf{r}) \phi^B(\mathbf{r}) \to 0$ for all $\mathbf{r}$. Consequently, the overlap contribution to the density vanishes identically, i.e., $\rho^{AB}(\mathbf{r}) = 0$.

The disappearance of the nonadditive terms requires a more subtle analysis. As the overlaps vanish, Eq.~(\ref{eqz3}) reduces to the form shown below, employing the notation defined in Eqs.~(\ref{eqz1}) and~(\ref{eqz2}):

\begin{gather}
\label{eq3}
S^{\text{add}}_{\rho,\infty}(\mathbf{r}) \to -\rho^{AA}(\mathbf{r}) \log \rho^{AA}(\mathbf{r}) - \rho^{BB}(\mathbf{r}) \log \rho^{BB}(\mathbf{r}) \\
\notag
S^{\text{nadd}}_{\rho,\infty}(\mathbf{r}) \to -\rho^{AA}(\mathbf{r}) \log \left\{ \frac{\rho^{AA}(\mathbf{r})}{\rho^{AA}(\mathbf{r}) + \rho^{BB}(\mathbf{r})} \right\} - \rho^{BB}(\mathbf{r}) \log \left\{ \frac{\rho^{BB}(\mathbf{r})}{\rho^{AA}(\mathbf{r}) + \rho^{BB}(\mathbf{r})} \right\}
\end{gather}

Since the centers are well separated, $\rho^{AA}({\bf r}) \gg \rho^{BB}({\bf r})$ for an ${\bf r}$ closer to $A$ than to $B$, and vice-versa. If we study the functions only for $\mathbf{r} \in \mathbb{R}^3 \setminus \{\mathbf{0}\}$ - real space divided into two regions by 
the equidistant plane $|{\bf r}|=0$ , which we denote by domain {\bf A} and {\bf B}), we have:
\eq{
\frac{\rho^{AA}({\bf r})}{\rho^{AA}({\bf r}) + \rho^{BB}({\bf r})} = \omega_{\text{{\bf A}}}({\bf r}) \ , \ \ \frac{\rho^{BB}({\bf r})}{\rho^{AA}({\bf r}) + \rho^{BB}({\bf r})} = \omega_{\text{{\bf B}}} ( {\bf r})
}
with $\omega_{\text{{\bf A}}}({\bf r}) = 1$ if ${\bf r} \in {\bf A}$, and $=0$ for ${\bf r}$ in ${\bf B}$; ie, the ratios are $1$ for ${\bf r}$ closer to $A$ (belonging to region {\bf A}) and $0$ if $\bf r$ belongs to region {\bf B} (and vice-versa). We thus have;
\eq{ \label{eq19}
- S^{nadd}_{\rho}({\bf r}) \to \ \rho^{AA}({\bf r}) \log \omega_{\text{{\bf A}}}({\bf r}) \ + \ \rho^{BB}({\bf r}) \log \omega_{\text{{\bf B}}}({\bf r}) 
}
Since $\log(1)=0$, when inside region {\bf A} the first term vanishes, but the second term also vanishes ($\lim_{x \to 0} x\log x=0$). Thus, it is proven that at very large distances between the centers, $S^{nadd}_{\rho} \to 0, \forall {\bf r}$ (the nonadditive entropy vanishes for all space). 

Thus, the correct expression for the infinite distance limit entropy is:
\eq{
\label{eq4}
S_{\rho, \infty} = \int S^{add}_{\rho, \infty}({\bf r}) d{\bf r} = \ - \int \rho^{AA}({\bf r}) \log \rho^{AA}({\bf r}) d{\bf r} \ - \  \int \rho^{BB}({\bf r}) \log \rho^{BB}({\bf r}) d{\bf r} = S_{\rho,\infty}^{A} + S_{\rho,\infty}^{B} = 2S_{\rho,\infty}^{A}
}
the last equality holding for homonuclear molecules. $S_{\rho,\infty}^{A}$ is the entropy of the isolated atom $A$, undeformed by the molecular environment. This is true for a minimal-basis approach. Beyond a minimal-basis, as we shall see, we may, in imperfect method/basis situations, have deformed atoms as we take the infinite distance limit of a molecule. 

If we turn our attention to the limits of the shape-function with two electrons  ($N=2$ does not introduce any loss of generality in the limits),  we can find after some simple algebra:
\begin{align} \label{eq21}
S_{\sigma} &= \int S^{add}_{\sigma}({\bf r}) d{\bf r} =
\frac{ S_{\rho,\infty}^{A}}{ N} + \frac{ S_{\rho,\infty}^{B}}{N } + \log N =  S_{\sigma,\infty}^A + \log 2, \notag
\end{align}
where $S_{\sigma,\infty}^A \equiv S_{\rho,\infty}^A /N_A$,  $N_A$ being the number of electrons of the isolated, unperturbed $A$ atom. 
The shape-function infinite distance limit is thus "contaminated" by a persisting $\log 2$ term. This electron number dependence introduces reference problems, for example, when trying to compare free atoms with atoms-in-the-molecules.
Since it is well-known \cite{matitojce} that the bond order in the ground-state Hartree-Fock wavefunction of H\textsubscript{2} does not vanish at infinite distance, the above expressions shows that, with a minimal basis, $S_\rho$ reduces to an unperturbed sum of two hydrogen atom entropies, with no indication of persisting bonding effects whatsoever. As we shall see next, this limit is independent, in practice, of the level of (static) correlation introduced, which is what allows for chemical bonding to vanish at large distances. Plots of  $S_\rho$  as a function of the internuclear distance, with the corresponding partition terms for the Hartree-Fock wavefunction, are included in Figure 1.


\subsubsection{The Heitler-London wavefunction}

The ground-state Heitler-London wavefunction for a two-electron homonuclear hydrogen molecule, minimally expressed in terms of two atomic 1$s$ orbitals, is:
\eq{
\psi_{\text{HL}}(\mathbf{r}_1, \mathbf{r}_2) = \frac{1}{\sqrt{2 + 2S^2}} \Big[ \phi^A(\mathbf{r}_1) \phi^B(\mathbf{r}_2) + \phi^B(\mathbf{r}_1) \phi^A(\mathbf{r}_2) \Big],
}
where $\phi^A$ and $\phi^B$ are atomic 1$s$ orbitals on different centers, with overlap integral $S$. This wavefunction correctly dissociates the molecule, since it sufficiently accounts for static correlation (see Figure S1 (in section 4) of Supplementary for a plot of the molecular energy as the internuclear distance is increased for the three different model wavefunctions examined in this section (HF, HL, and FCI)).
The electron density for this wavefunction can be expressed as:
\eq{
\rho (\mathbf{r}) = \frac{1}{1+S^2} \left[  {\phi_A^2(\mathbf{r}) + \phi_B^2(\mathbf{r}) + 2S \phi^A(\mathbf{r}) \phi^B(\mathbf{r})} \right], 
}
so that, employing the previously developed partitioning procedure, we  can write:
\begin{gather}
\label{eq12} 
S^{add}_{\rho} ({\bf r}) = \ - \Gamma^{AA}({\bf r}) \log \Gamma^{AA}({\bf r}) \ - \ \Gamma^{BB}({\bf r}) \log \Gamma^{BB}({\bf r})  - 2S\Gamma^{AB}({\bf r}) \log \Gamma^{AB}({\bf r})\\
\notag S^{nadd}_{\rho}({\bf r}) = \ \Gamma^{AA}({\bf r}) \log \Bigg\{ \frac{\Gamma^{AA}({\bf r})}{\Gamma^{AA}({\bf r}) + \Gamma^{BB}({\bf r}) + 2S\  \Gamma^{AB}({\bf r})} \Bigg\} \\
\notag + \ \Gamma^{BB}({\bf r}) \log \Bigg\{ \frac{\Gamma^{BB}({\bf r})}{\Gamma^{AA}({\bf r}) + \Gamma^{BB}({\bf r}) + 2S\  \Gamma^{AB}({\bf r})} \Bigg\} \\
\notag + \ 2S\ \Gamma^{AB}({\bf r}) \log \Bigg\{ \frac{\Gamma^{AB}({\bf r})}{\Gamma^{AA}({\bf r}) + \Gamma^{BB}({\bf r}) + 2S\  \Gamma^{AB}({\bf r})} \Bigg\},
\end{gather}
with $\Gamma^{\mtc{AB}} \equiv \frac{\phi^{\mtc{A}} \phi^{\mtc{B}}}{1+S^2}$. 
As in the previous subsection, the infinite distance limits of this object emerge as purely additive atomic contributions
\eq{
S_{\rho}=2S^A_{\rho} \ \ \text{or} \ \ S_{\sigma}=S^A_{\sigma} + \log2 
}
for the electron-density and shape-function, respectively, just as in the HF case. This stems from the fact that the minimal basis set densities at dissociation are equal in both models, revealing that the spurious bonding effects present in the HF wavefunction as the molecule dissociates do not propagate into a distinct behaviour in $S_\rho$. Actually, only the rate at which the entropy achieves its limit and the presence of an internuclear distance window  where $S_\rho > 2S^A_\rho $  in the HF case, provide any qualitative difference between them.  
%

\subsubsection{The multideterminant case}

The electron density for multideterminant wavefunctions of a (heteronuclear) diatomic molecule can generally be written in terms of the atomic orbitals as:
\eq{
\label{eq14}
\rho ({\bf r})  = \sum_{\mtc{A}=A}^{B} \sum_{i\in \mtc{A}} \sum_{\mtc{B}=A}^{B} \sum_{j\in \mtc{B}} C_{ij}^{\mtc{AB}} \phi_i^{\mtc{A}} ({\bf r})  \phi_j^{\mtc{B}} ({\bf r}) 
}
where $C^{\mtc{AB}}_{ij}$ are coupling coefficients that  incorporate the determinant coefficients and the relevant overlaps. Since this is for diatomic molecules, the sums on $\mtc{A}$ and $\mtc{B}$ are over the two distinct centers.
Our previous approach applies equally in this case, provided we define: 
\eq{
\rho^{\mtc{A}\mtc{B}} ({\bf r}) \equiv 
\sum_{i\in \mtc{A}} \sum_{j \in \mtc{B}} C^{\mtc{AB}}_{ij} \phi^{\mtc{A}}_i ({\bf r})\phi^{\mtc{B}}_j ({\bf r}),
}
$i$ and $j$ summing over all atomic orbitals restricted to centers $\mtc{A}$ and $\mtc{B}$, respectively. The partition of the entropy into additive and non-additive contributions for H\textsubscript{2} holds exactly as in Equation \ref{eq3}. 
Provided that this model wavefunction, which is equivalent to a minimal basis CAS(2,2)\footnote{We denote a Complete Active Space calculation with $n$ electrons and $m$ unique spatial orbitals as CAS(n,m)} (and for H\textsubscript{2} is a FCI) calculation does also dissociate correctly, it is straightforward to show (see the SI section 3 for further details) that the entropies tend to exactly the same limits already examined.

\begin{figure}[h!]
    \centering
    \includegraphics[width=0.7\textwidth]{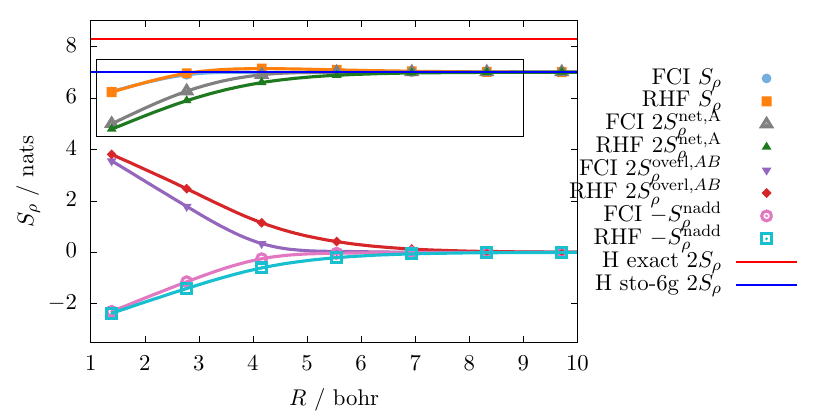}
    \includegraphics[width=0.7\textwidth]{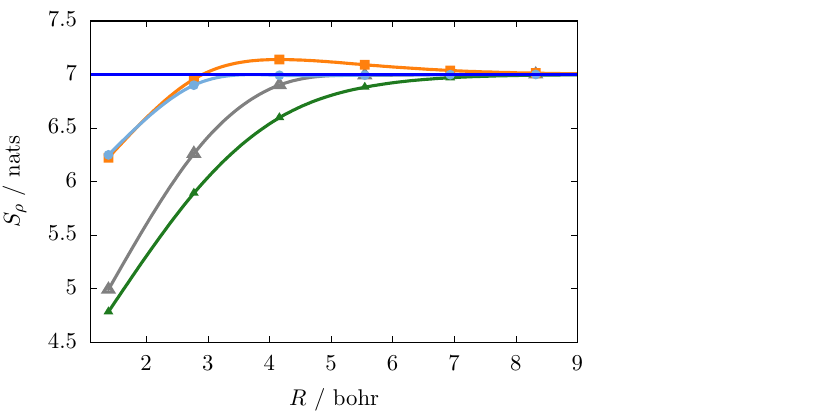}
    \caption{Various entropic terms for H\textsubscript{2} described with a minimal STO-6G basis for different methods at various internuclear distances (points). The sum of the free atomic entropies obtained with the same basis set is shown as a straight horizontal line. Note the convergence of all methods to solely net contributions that equal the isolated atom entropy values. The lowest $R$-value is the equilibrium internuclear distance. The bottom plot corresponds to the highlighted region. We  plot the negative of the nonadditive entropy since the nonadditive part of  $S_\rho$ is $-S^{nadd}$.}
    \label{fig:h2-entropies-minimal}
\end{figure}

The calculated entropy terms (full, net atomic, total overlap, nonadditive) at various internuclear distances for H\textsubscript{2} (FCI versus HF) and N\textsubscript{2} (CAS(6,6) versus HF) molecules, as computed with a STO-6G minimal basis set, are included in Figures \ref{fig:h2-entropies-minimal} and \ref{fig:n2-entropies-minimal}, respectively. They fully support the algebraic limits and considerations of the previous subsections. Since in both systems the limiting densities collapse into the sum of two non-interacting free atoms densities,  the  only contributing entropic term at this limit is the net entropy. This is true for the three types of wavefunctions. As noted, HF is the only method with a full entropy (orange line) that does not increase monotonically towards the infinite distance limits, becoming larger than that for distances $R\gtrapprox 3 \text{\ bohr}$ (H\textsubscript{2}) and $R\gtrapprox 3.5 \text{\ bohr}$ (N\textsubscript{2}). This is clearly due to the presence of overlap and nonadditive contributions, since the HF atomic net entropy (in green) does not suffer from this problem. We have found that this is basically the only qualitative indicator of the wrong dissociative behavior of the HF wavefunction that is  contained in $S_\rho$.


\begin{figure}[h!]
    \centering
    \includegraphics[width=0.8\textwidth]{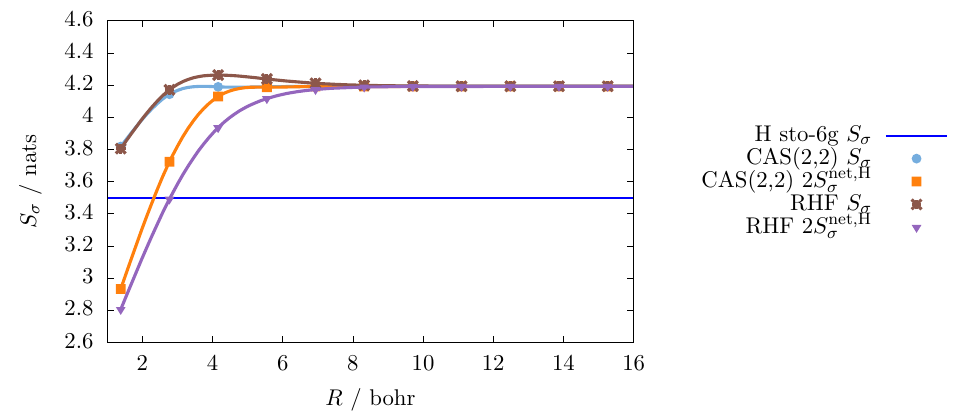}
    \caption{Shape-function net and total entropies for STO-6G H\textsubscript{2}  as obtained from different methods at various internuclear distances (points). The entropy of the isolated atoms is depicted with a straight horizontal line. The lowest $R$-value is the equilibrium internuclear distance. Notice the non-extensivity of $S_\sigma$.).
    }
    \label{fig:h2-shape-minimal}
\end{figure}

\begin{figure}[h!]
    \centering
    \includegraphics[width=0.8\textwidth]{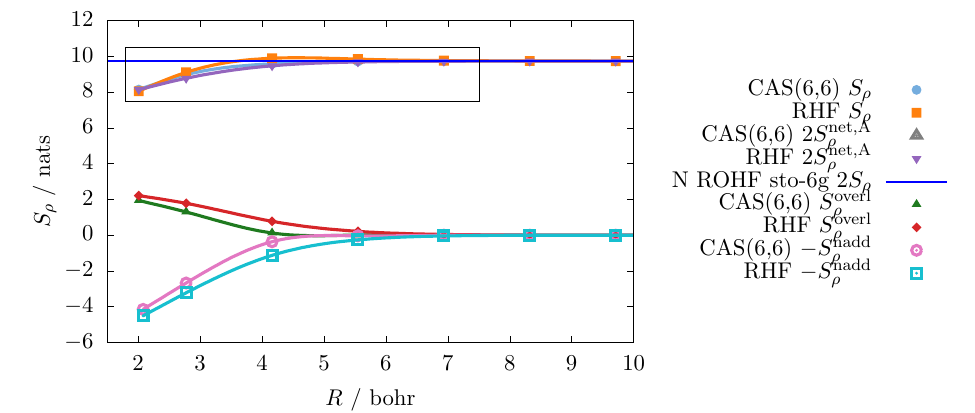}
    \includegraphics[width=0.8\textwidth]{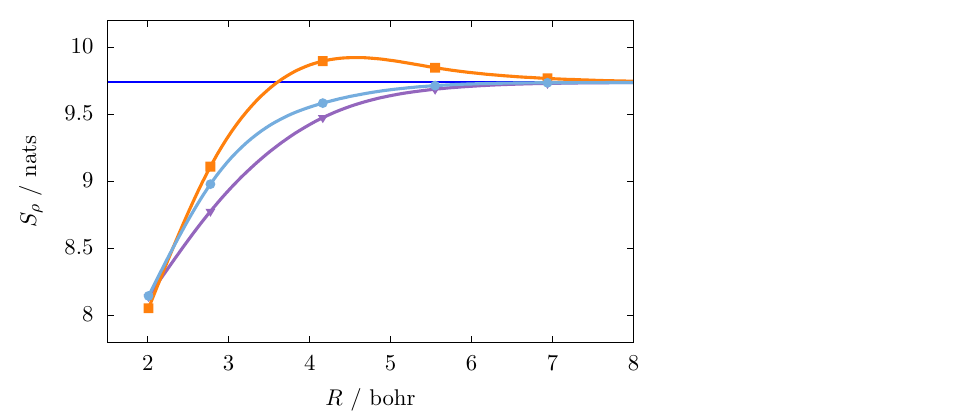}
    \caption{Various entropic terms for N\textsubscript{2} described with a minimal STO-6G basis for different methods at various internuclear distances (points). The sum of the free atomic entropies obtained with the same basis set is shown as a straight horizontal line. Note the convergence of all methods to solely net contributions that equal the isolated atom entropy values. The lowest $R$-value is the equilibrium internuclear distance. The bottom plot corresponds to the highlighted region. We  plot the negative of the nonadditive entropy since the nonadditive part of  $S_\rho$ is $-S^{nadd}$.)}
    \label{fig:n2-entropies-minimal}
\end{figure}

\begin{figure}[h!]
    \centering
    \includegraphics[width=0.8\textwidth]{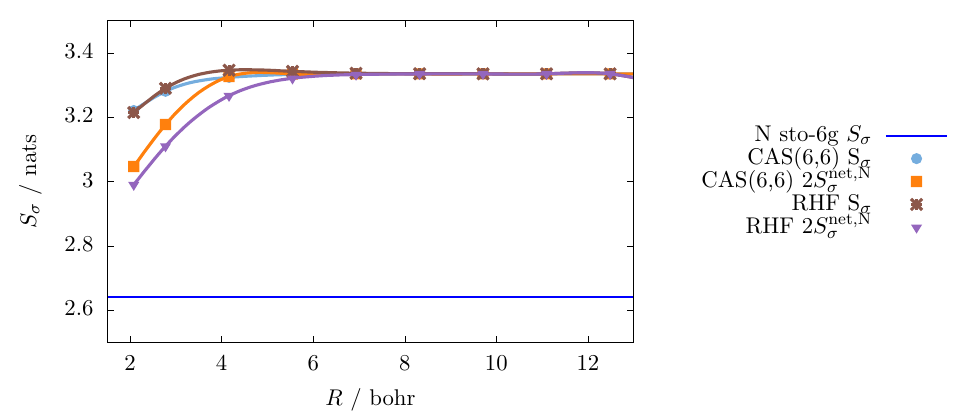}
    \caption{Net and total entropies (shape-function based) terms for N\textsubscript{2} at minimal-basis (STO-6G) for different methods at various internuclear distances (points) and entropy of the isolated atoms (lines). The lowest x-value is the equilibrium internuclear distance. Notice how the values at large distances are close in magnitude to the values of the H\textsubscript{2} molecule, which is an advantage of using shape-function based entropies. Similarly to H\textsubscript{2}, the atomic limits are not obeyed. The predicted difference between the full entropy and the atomic shape-function entropy is $\log N - \log{N_A} = \log(2) = 0.69315 \text{\ nats}$, matching the numerical difference. 
    }
    \label{fig:n2-shape-minimal}
\end{figure}

In terms of the shape-function integrations, Figures \ref{fig:h2-shape-minimal} (H\textsubscript{2}) and \ref{fig:n2-shape-minimal} (N\textsubscript{2}) show that convergence to the entropy of two isolated atoms no longer holds. If we use the atomic shape-function (net) entropy as a reference, we see that for both molecules the total integrated shape-function entropy limit obeys $S^{\infty}_\sigma = S^{\text{atom}}_\sigma + \log 2$.

In the next section, we demonstrate that the above results are actually independent of the system's specifics. We also analyze the case for extended basis-sets, where similar conclusions are drawn.

\subsection{The dissociation limit of the Shannon entropy}

Since the following  expression for the density partition is universally valid (we now drop the implicit summation superindex),  %
\eq{
\label{eqgen1}
\rho ({\bf r})  = \sum_{\mtc{A}} \sum_{\mtc{B}} \rho^{\mtc{A B}} ({\bf r}),
}
one can deduce general properties for the behavior of the Shannon entropy %
in the limit of infinite distance between all pairs of centers, $R= |{\bf R}_\mtc{A}  - {\bf R}_\mtc{B}| \to \infty \ \ \forall \{\mtc{A, B \neq A}\}$. %
In  this limit, the overlap entropies vanish for all atomic pairs, %
since 
$  \phi^{\mathcal{A}}_i({\bf r}) \phi^{\mathcal{B}}_j({\bf r}) \to 0 \quad \forall \, i \in \mathcal{A}, \, j \in \mathcal{B} \ (\mathcal{A} \neq \mathcal{B}), \, \forall \, {\bf r} 
\implies \rho^{\mathcal{AB}}({\bf r}) = \rho^{\mathcal{BA}}({\bf r}) = 0 \quad \forall \, \mathcal{A} \neq \mathcal{B}$.

Then, as shown in previous section, when the overlap terms vanish, so does the nonadditive entropy  (Equation (\ref{eq19})), and the Shannon entropy becomes effectively additive into its atomic contributions. 
This was shown by the natural splitting of the spatially-resolved diatomic electron density into two non-overlapping regions {\bf A} and {\bf B} when their respective centers are far apart. The splitting,  in turn, decouples the entropy integral in molecular space into two separate integrals, one over each region:
\eq{
\int \rho({\bf r}) \log \rho ({\bf r}) d{\bf r}  \to \int_{\text{A}}  \rho({\bf r}) \log \rho ({\bf r}) d{\bf r} + \int_{\text{B}}  \rho({\bf r}) \log \rho ({\bf r}) d{\bf r}, 
}
so that the entropy  becomes additive over those regions, which contain a single atom.   %
These conclusions  extend to molecules of any number of atoms 
so that, when $\phi^{\mtc{A}}_i({\bf r}) \phi^{\mtc{B}}_j({\bf r}) \to 0$, 
\eq{
\int \rho({\bf r}) \log \rho ({\bf r}) d{\bf r} \to \sum_{\mtc{{\bf A}}} \int  \rho({\bf r}) \log \rho ({\bf r}) d{\bf r} 
}
and $S^{nadd}=0$.  As we will see, this result is not valid for the Rényi entropies in their different forms.
With these considerations, the following expression is generally valid: 
\eq{
S_{\rho, \infty} = \int S^{add}_{\rho,\infty}({\bf r}) d{\bf r} =  \sum_{\mtc{A}} S^{add, \mtc{A}}_{\rho,\infty},
}
where
$S^{add, \mtc{A}}_{\rho,\infty}= -\int \rho_\infty^{\mtc{AA}}({\bf r}) \log \rho_\infty^{\mtc{AA}} ({\bf r}) d{\bf r}$ is 
the $\mtc{AA}$ center contribution at infinite separation (which is also the only contribution due to atom $\mtc{A}$ in this limit). For the shape function we have instead:
\eq{
S_{\sigma, \infty} = \int S^{add}_{\sigma,\infty}({\bf r}) d{\bf r} =  \sum_{\mtc{A}} S^{add, \mtc{A}}_{\sigma,\infty}  = \sum_{\mtc{A}} \frac{N_\mtc{A}}{N} S^\mtc{A}_\sigma
 - \sum_{\mtc{A}} \frac{N_\mtc{A}}{N}   \log N_\mtc{A} + \log N
 }
since here $S^{add, \mtc{A}}_{\sigma,\infty}= -\int \left[\rho_\infty^{\mtc{AA}}({\bf r}) /N\right] \log \left[\rho_\infty^{\mtc{AA}} ({\bf r}) /N\right] d{\bf r}  $ and the reference  atomic entropy is $S^{\mtc{A}}_{\sigma}= -\int \rho_\infty^{\mtc{AA}}({\bf r}) \log \rho_\infty^{\mtc{AA}} ({\bf r}) 
d{\bf r}= -\int \left[\rho_\infty^{\mtc{AA}} ({\bf r}) /N_\mtc{A}\right] \log \left[\rho_\infty^{\mtc{AA}} ({\bf r}) /N_\mtc{A}\right]  d{\bf r}$.

Now, in a minimal-basis set $\rho_\infty^{\mtc{AA}}$ is typically constrained to coincide with the free atomic density, and  the total molecular entropy will tend to the sum of its free atomic values.  Beyond a minimal-basis,  however (see the SI section 5),  $\rho_\infty^{\mtc{AA}}$ needs not coincide with the isolated-atom limit built with the same basis set. Actually, the  $\rho_\infty^{\mtc{AA}}$ limit is in general a deformed atomic density, meaning that %
\eq{
S_{\rho, \infty} = \int S^{add}_{\rho,\infty}({\bf r}) d{\bf r} = \sum_{\mtc{A}} S_{\rho,\text{def}}^{\mtc{A}},
}
$S_{\rho,\text{def.}}^{\mtc{A}}$ being the deformed-density entropy for the isolated atom $\mtc{A}$.
Similarly,  for the shape-function:
\eq{
S_{\sigma, \infty} = \int S^{add}_{\sigma,\infty}({\bf r}) d{\bf r} = \sum_{\mtc{A}} \frac{N_A}{N} S_{\sigma,\text{def.}}^{\mtc{A}}  
- \sum_{\mtc{A}} \frac{N_\mtc{A}}{N} \log \frac{N_\mtc{A}}{N}.
 }
Notice that $N_A$ will be typically, but not necessarily, equal to the number of electrons of the dissociating atom $A$.  The factor $- \sum_{\mtc{A}}^\mtc{N} \frac{N_\mtc{A}}{N} \log \frac{N_\mtc{A}}{N}$ for homonuclear molecules that dissociate into neutral atomic densities reduces to $\log{\mtc{N}}$, where $\mtc{N}$ is the number of centers. For homonuclear diatomics this is simply $\log(2)$, as found in Figures \ref{fig:h2-shape-minimal}, \ref{fig:n2-shape-minimal}.
%

\begin{figure}[h!]
    \centering
    \includegraphics[width=0.8\textwidth]{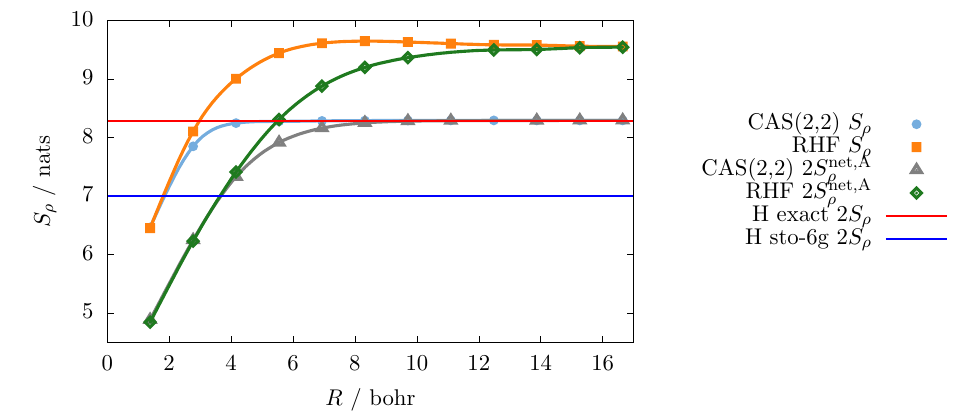} 
    \includegraphics[width=0.8\textwidth]{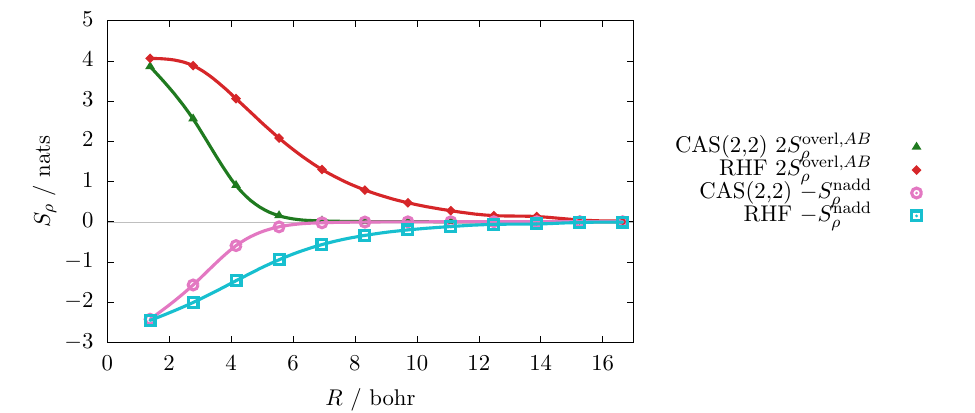}
 \caption{
     H\textsubscript{2}/aug-cc-pVTZ at the HF and CAS(2,2) levels at different internuclear distances. Top:  $S_\rho$ and $S_\rho^{net}$  entropies. The sum of the entropies of the two H/aug-cc-pVTZ isolated atoms is shown as a horizontal line, and the lowest $R$-value is the equilibrium internuclear distance. Bottom: $S_\rho^{overl}$ and $-S_\rho^{nadd}$
}
    \label{fig:h2-net-full-aug}
\end{figure}
\begin{figure}[h!]
    \centering
    \includegraphics[width=0.8\textwidth]{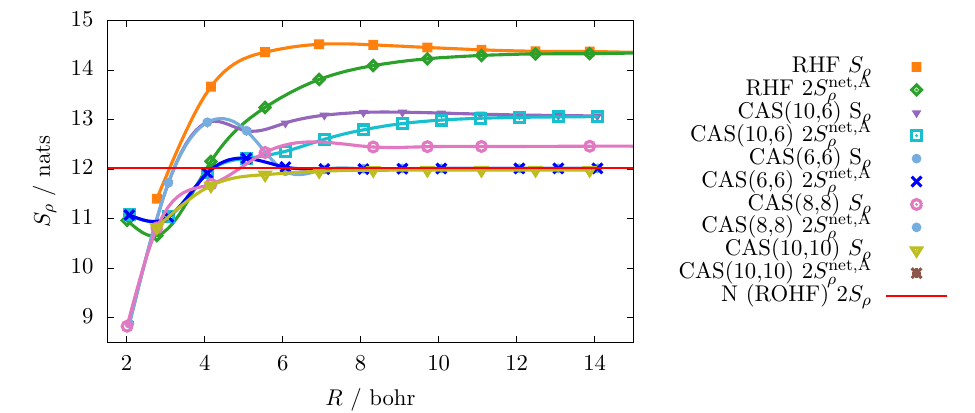}
    \includegraphics[width=0.8\textwidth]{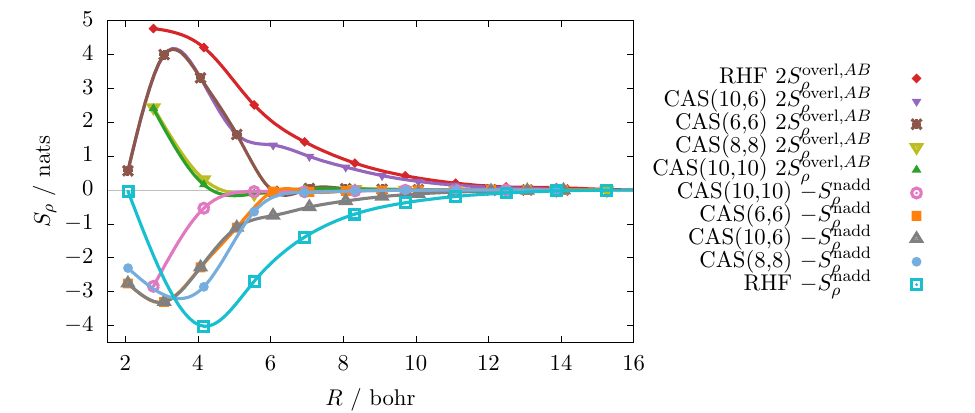}
    \caption{
    N\textsubscript{2}/aug-cc-pVTZ at the HF and several CAS levels at different internuclear distances. Top:  $S_\rho$ and $S_\rho^{net}$  entropies. The sum of the entropies of the two ROHF/aug-cc-pVTZ isolated N atoms is shown as a horizontal line, and the lowest $R$-value is the equilibrium internuclear distance. Bottom: $S_\rho^{overl}$ and $-S_\rho^{nadd}$. Only the full-valence-CAS(10,10) calculation is size extensive and $S_\rho$ for the isolated N atom is $6.00001$ and $6.00800$ nats at the HF and CAS(7,5)/aug-cc-pVTZ levels, respectively. 
    }
    \label{fig:n2-net-full-aug}
\end{figure}
\begin{figure}[h!]
    \centering
    \includegraphics[width=0.8\textwidth]{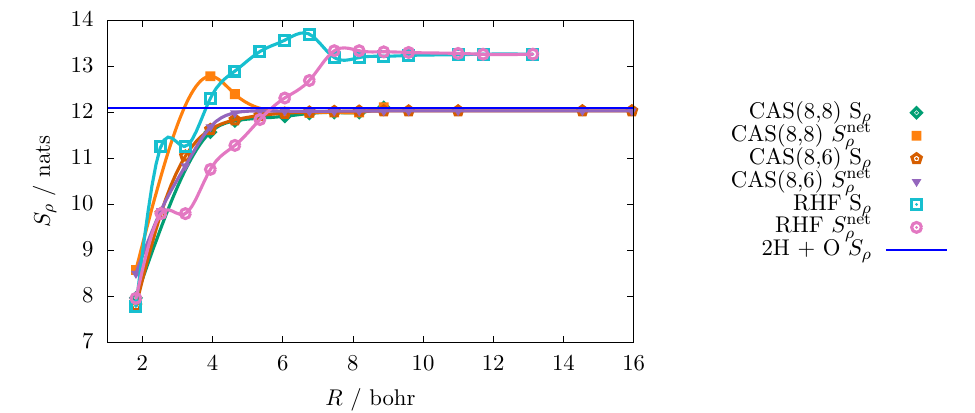}
    \includegraphics[width=0.8\textwidth]{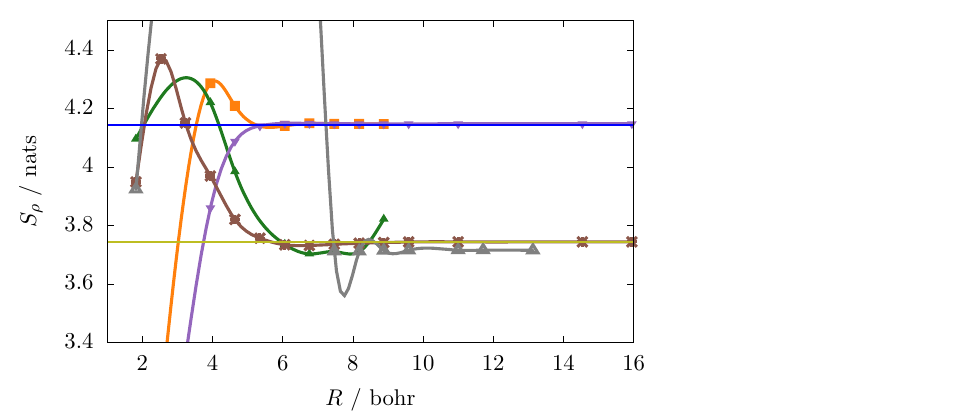}
    \caption{
    Total and net entropies for H\textsubscript{2}O at aug-cc-pVTZ for different methods at various internuclear distances (points) and entropy of the isolated atoms (lines). The internuclear distances are the O-H distance, which is simultaneously changed for both bonds, preserving the symmetry of the molecule during dissociation. The lowest x-value is the equilibrium internuclear distance. Notice how CAS(8,6) smoothly tends to the correct atomic limit for both H and O atoms, but HF behaves erratically. The lower figure is focused solely on the net atomic entropies and their limits, for resolution. The atomic limits converge smoothly to the H atom for these distances, but not for the O atom. Atomic limits are for ROHF.
    }
    \label{fig:h2o-net-full-aug}
\end{figure}
\begin{figure}[h!]
    \centering
    \includegraphics[width=0.8\textwidth]{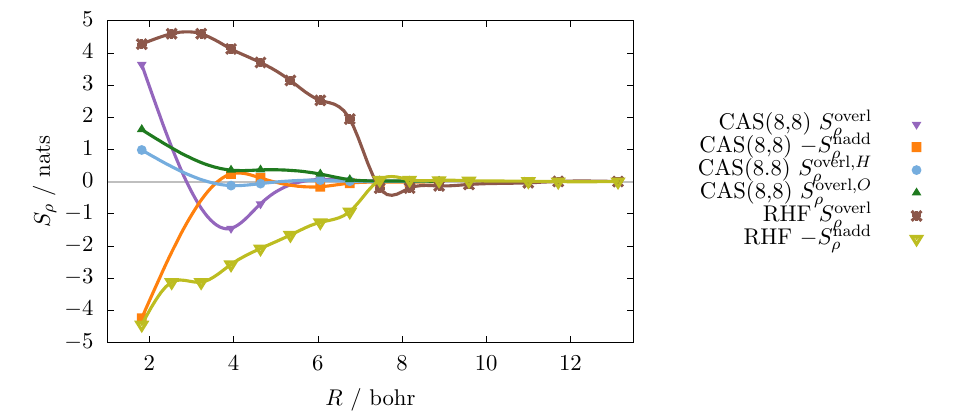}
    \caption{Overlap and nonadditive entropies for H\textsubscript{2}O at aug-cc-pVTZ for different methods at various internuclear distances (points) and entropy of the isolated atoms (lines). The lowest x-value is the equilibrium internuclear distance. Notice the vanishing of the overlap and nonadditive entropies for all methods. $S^{overl}=S^{overl,\text{O}} + 2S^{overl,\text{H}} = \sum_A S^{overl,\text{OA}} + \sum_B S^{overl,\text{HB}}$.  (We chose to plot the negative of the nonadditive so it becomes clearer how it enters the overall entropy, since the nonadditive part is $-S^{nadd}$.)}
    \label{fig:h2o-nadd-overlap-aug}
\end{figure}

Numerical results for the different entropic terms built with an aug-cc-pVTZ basis that support these conclusions are found in Figure \ref{fig:h2-net-full-aug} for H\textsubscript{2}, Figure \ref{fig:n2-net-full-aug} for N\textsubscript{2}, and Figures \ref{fig:h2o-net-full-aug} and  \ref{fig:h2o-nadd-overlap-aug} for H\textsubscript{2}O. 
A first interesting discussion appears in the HF description of H\textsubscript{2}.  Due to the incorrect dissociative behavior of the Hartree-Fock approximation, which introduces spurious ionic contributions at any internuclear distance, the limiting atomic densities $\rho_\infty^{\mtc{AA}}$ do not coincide with aug-cc-pVTZ free H atoms, and the entropy saturates at large interatomic distances to a value that does not coincide with twice the H atomic entropy.  We find that the entropy of an HF calculation with an extended basis tends to a considerably higher value than that found for a correctly dissociating CAS(2,2) calculation, where the inclusion of static correlation effects guarantees that $S_\rho$ tends exactly to $2\times S_{\rho,\text{H}}$. This incorrect limit is entirely determined by the incorrect $S^{net}$ behaviour. 

Another relevant consideration regards the non-net contributions to $S_\rho$ that can be found in the bottom panel of Figure~\ref{fig:h2-net-full-aug}. They are considerably larger in the HF approximation,  and  close to their CAS counterparts at distances close to equilibrium. Actually, at equilibrium, where HF is a reasonable approximation to the dihydrogen wavefunction, the values of $S_\rho$ in both approximations are similar, pointing to $S^{nadd}$ as a possible measure of bonding effects in molecules that should be explored further.  

A similar analysis in dinitrogen, see Figure \ref{fig:n2-net-full-aug}, reinforces our previous results. We have considered several CAS calculations with different active spaces, from a small CAS(6,6) to a full-valence CAS(10,10), as well as the base-level HF wavefunction.  Notice that only the full-valence CAS dissociates correctly to two ground-state N atoms, while the smaller active spaces and HF allow for the inclusion of a limited or null amount of static correlation, respectively.  Plots of the evolution of the total molecular energy can be found in the SI, Figures S3 and S4. It is clear that the limit of $S_\rho$ is determined by its net contribution, that describes deformed N atoms except in the full-valence CAS calculation. In agreement with previous insights, the deformation is largest at the HF level. 

Finally, a symmetric dissociation of a water molecule in which a simultaneous stretching of the two O-H bonds preserving the H-O-H angle and the $C_{2v}$ symmetry has been examined.  As in previous examples, both HF and several CAS wavefunctions differing in the active spaces have been selected, and the results are shown in Figures \ref{fig:h2o-net-full-aug} and  \ref{fig:h2o-nadd-overlap-aug}. 
We find that, again, only full-valence CAS(8,8) correctly contains (approximately) the atomic limit of the entropy (see also the energy plots in Figure S5) at large distances.  While the H atom contribution limit clearly integrates to the isolated H atom entropy, such a limit is only approximately fulfilled in the case of the O atom. It is well known that CAS calculations are difficult to maintain in a smooth electronic state upon dissociation. Nevertheless, the oxygen's limit is close to the ROHF reference.
Thus, the entropy behaves as expected for the full-valence CAS. Similar trends  are found for HF and CAS(8,6) cases, although it becomes increasingly difficult to correctly dissociate the molecule and obtain smooth entropy and energy curves. In this latter case, the CAS(8,6) calculation converges to the atomic limit of oxygen better, but this might well be a numerical coincidence. In SI Figures S7 and S8, we provide plots of the atomic-resolved components of the entropy for further information. By the analysis of the behavior of the energies on dissociation, we confirmed the dissociation of the O atom into its triplet state.

\section{Rényi entropies of the electron density}

\subsection{Partition into atom-pairs of the Rényi entropy for general $\alpha$}

The Rényi entropy is a generalization of the Shannon entropy where one relaxes the fourth Kinchin axiom \cite{kinchin} of strong additiveness for independent probabilities.
Its form is dependent on the choice of a parameter $\alpha > 0$. For a set of events $\{i\}$ with probabilities $p_i$, 
\eq{
S^{\alpha} = \frac{1}{1-\alpha} \log \Big\{ \sum_i (p_i)^{\alpha} \Big\} 
} 
is called the Rényi entropy of order $\alpha$. The choice of $\alpha$ affects the weighting of the different probabilities $p_i$. In the $\alpha=0$ case, all states are equally weighted, while higher values of $\alpha$ will give higher weighting to larger probability contributions, with the $\alpha \to \infty $ limit converging to only the largest probability (or probabilities if degenerate) contributing to the entropy. For $\alpha \to 1$ one recovers the Shannon entropy, as can be shown by using L'Hopital's rule, and when $\alpha=2$ it is sometimes known as the \textit{collisional entropy} or the \textit{(Rényi) entanglement entropy}.

In the general $\alpha$ case, the von Neumann-like form of the Rényi entropy  applied to a density matrix $\hat\rho$ is:
\eq{
S_{vN}^{\alpha} = \frac{1}{1-\alpha} \log \Big\{ \trace{ \hat \rho^{\alpha} } \Big\} 
} 
where $\hat \rho^{\alpha}$ naturally involves matrix multiplication.
For the electron-density (or shape-function), the set of Rényi entropies can be expressed as:
\eq{
S^{\alpha}_{\rho} = \frac{1}{1-\alpha} \log \Big( \int \rho({\bf r})^{\alpha} d{\bf r} \Big) 
}
(for the shape function, one divides the density by $N^\alpha$). We will again use $\rho$ as a proxy for either the electron-density or the shape-function $\rho/N$. 
Since this entropy is not generally of trace-form, one cannot obtain its value for a single value of {\bf r}, but only its total integrated value. This means that unless $\alpha=1$ (the Shannon entropy case previously handled) one cannot define an entropy density, and thus we now work solely with the total entropy values.

Similarly to what was previously done, the electron density can be expressed in terms of atomic orbitals or primitive functions  $\{ \phi \}$ and atomic centers $\mtc{A}$ . Using the density decomposition found in  Equation (\ref{eqx2}), %
we partition the generalized Rényi entropy for $\alpha = 2$ either using the density or the shape function.  Details are given in SI section 6. In the following,  we define a 4-index object $p^{\mtc{AB}}_{\mtc{CD}} = p^{\mtc{CD}}_{\mtc{AB}}  \equiv \frac{\int \rho^{\mtc{AB}} ({\bf r}) \rho^{\mtc{CD}} ({\bf r}) d{\bf r}}{ \int \rho ({\bf r})^2 d{\bf r}} = \frac{\int \sigma^{\mtc{AB}} ({\bf r}) \sigma^{\mtc{CD}} ({\bf r}) d{\bf r}}{ \int \sigma ({\bf r})^2 d{\bf r}}$, so that the additive and non-additive Rényi entropies become:
\begin{align}
S^{\text{add}, \alpha=2}_{\rho} = - \sum_{\mtc{ABCD}} p^{\mtc{AB}}_{\mtc{CD}} \log \int \rho^{\mtc{AB}}({\bf r}) \rho^{\mtc{CD}}({\bf r}) d{\bf r} \\
S^{\text{nadd}, \alpha=2}_{\rho} = \sum_{\mtc{ABCD}} p^{\mtc{AB}}_{\mtc{CD}} \log p^{\mtc{AB}}_{\mtc{CD}}
\end{align}
because of the order of $\alpha=2$ we must now sum over four center indices, $\mtc{A, B, C, D}$.

\subsection{General infinite distance limits of the Rényi entropy - persisting nonadditivity}

We turn our attention to the general molecule case (not restricted to two-centers) $\alpha=2$ Rényi entropy in the infinite distance limit, as previously analyzed in the Shannon ($\alpha=1$) case.
As for the Shannon entropy, all additive and nonadditive terms that have different orbital center indices vanish at infinite distance. This means that only the $p^{\mtc{AA}}_{\mtc{AA}}$ terms will survive in the infinite distance 2nd order Rényi entropy. Mimicking our previous manipulations,  the full entropy becomes a weighted sum of additive atomic entropies and a residual nonadditive entropy. If we define $p_\mtc{A} \equiv p^{\mtc{AA}}_{\mtc{AA}}$, we have:
\eq{ \label{eqx}
S^{\alpha=2}_{\rho, \infty} = - \sum_{\mtc{A}} p_{\mtc{A}} \log \int \rho^{\mtc{AA}}({\bf r})^2 d{\bf r} \ - \ \Big( - \sum_{\mtc{A}} p_{\mtc{A}} \log p_{\mtc{A}} \Big) = S^{\alpha=2,\text{net}}_{\sigma} \ - \ S^{\alpha=2,\text{nadd}}_{\sigma} = S^{\alpha=2}_{\rho,\infty} \ + \ 2\log N, 
}
\eq{
S^{\alpha=2}_{\rho,\infty} = - \sum_{\mtc{A}} p_{\mtc{A}} \log \int \rho^{\mtc{AA}}({\bf r})^2 d{\bf r} \ + \ \sum_{\mtc{A}} p_{\mtc{A}} \log p_{\mtc{A}}
}
This is valid for any number of centers. Using the general definition $p^\mtc{A}_{\alpha} \equiv \frac{\int \rho^{\mtc{AA}} ({\bf r})^\alpha d{\bf r}}{ \int \rho ({\bf r})^\alpha d{\bf r}} $, the following expressions are valid at any distance and for any order Rényi entropy:
\begin{align}
S^{\text{net},\alpha}_{\rho}= \sum_{\mtc{A}} p^A_{\alpha} S^{\text{net},\mtc{A},\alpha}_{\rho} = {\frac{1}{1-\alpha}} \sum_{\mtc{A}} p_{\alpha}^\mtc{A} \log \int  \rho^{\mtc{AA}} ({\bf r})^\alpha d{\bf r}, \\
S^{\text{nadd},\alpha}_{\text{intra}}= \sum_{\mtc{A}} S^{\text{nadd},\mtc{A},\alpha}_{\text{intra}} = {\frac{1}{1-\alpha}} \sum_{\mtc{A}} p_{\alpha}^\mtc{A} \log p_{\alpha}^\mtc{A}.
\end{align}
$S^{\text{nadd},\alpha}_{\text{intra}}$ refers only to the intra-atomic part (same center indices) of the nonadditive entropy for general distances. The nonadditive contribution is independent in form whether one is using the shape-function or electron-density. The net contribution, $S^{\text{net},\alpha}_{\rho}$, is equal to the additive contribution in the infinite distance limit.
These are not, of course, the only terms contributing to the entropy at non-infinite distances, but only those that will be relevant to analyze as we vary the distance and go into the domain of very large distances.

The presence of the nonadditive contribution and the weighting of the atomic entropies at infinite distance also shows an extensivity problem with the Rényi entropy even in its electron-density-based form. This results from the unequal weighting of low and high probabilities,  which is dependent on the value of $\alpha$. The higher the value of $\alpha$, the smaller this persisting nonadditive contribution becomes.
A comparison of the Rényi entropy of atoms in different molecular environments is thus more difficult than in the Shannon entropy case, and although Rényi entropies are very interesting mathematical constructs they present problems when we try to endow them with a chemical interpretation.

Take, for example, a homonuclear diatomic molecule. Then:
\eq{
S^{\alpha,\infty}_\rho = \frac{1}{1-\alpha} \log \int \rho^\mtc{AA} ({\bf r}) d{\bf r} - \frac{1}{1-\alpha} \log 2 = S^{\text{net},\alpha,\mtc{A}} + \frac{1}{1-\alpha} \log 2
}
which shows that it fails to describe any of the correct atomic  (assuming $S^{\text{net},\alpha,\mtc{A}}$ is the correct atomic $\alpha$-entropy) or deformed atomic limit. In the case of the shape-function, this problem adds to its already discussed non-extensivity features. We will formulate this precisely through equations below.

This additional nonadditive contribution is obviously present even in the minimal-basis H\textsubscript{2} example, appearing independently of the presence of static correlation. As we have established this is only dependent on the limit of the density itself, and since the entropic forms are mathematical functionals of this electron density, they obey similar limiting behaviors.

Concerning the relationship to atomic limits, we see a picture similar to that found in the case of Shannon entropy, with the necessary generalizations.
\eq{
S^{\alpha}_{\rho, \infty} = S^{add,\alpha}_{\rho,\infty}= \sum_{\mtc{A}} p_\alpha^\mtc{A} S^{\alpha,\mtc{A}}_{\rho,\text{def.}} - \frac{1}{1-\alpha} \sum_{\mtc{A}} p_\mtc{A} \log p_\mtc{A}
}
\eq{
S^{\alpha}_{\sigma, \infty} = \sum_{\mtc{A}} p_\alpha^\mtc{A} S^{\alpha,\mtc{A}}_{\sigma,\text{def.}} - \frac{1}{1-\alpha} \sum_{\mtc{A}} p^\alpha_\mtc{A} \log p_\mtc{A}  + \frac{\alpha}{1-\alpha} \sum_{\mtc{A}} p^\alpha_\mtc{A}  \log \frac{N_A}{N},
}
valid for any method/basis and molecule (a minimal-basis should return the pure, undeformed atoms), and taking into account $S_{\sigma,\text{def}}^{\alpha, \mtc{A}}$ divides the electron density by $N_A$ electrons when computing the shape function.
Likewise, we assume that within the method/basis, the dissociation limit exists for all atoms and that it yields atoms and not ions. If ions are generated, one must use $N_\mtc{A}$ for the ionic species, and the expected entropy of the atoms in both previous equations will now be the entropy of the free ions.
As can be seen, extensivity problems of these entropies are only made worse by considering the general Rényi picture.

\begin{figure}[h!]
    \centering
    \includegraphics[width=0.8\textwidth]{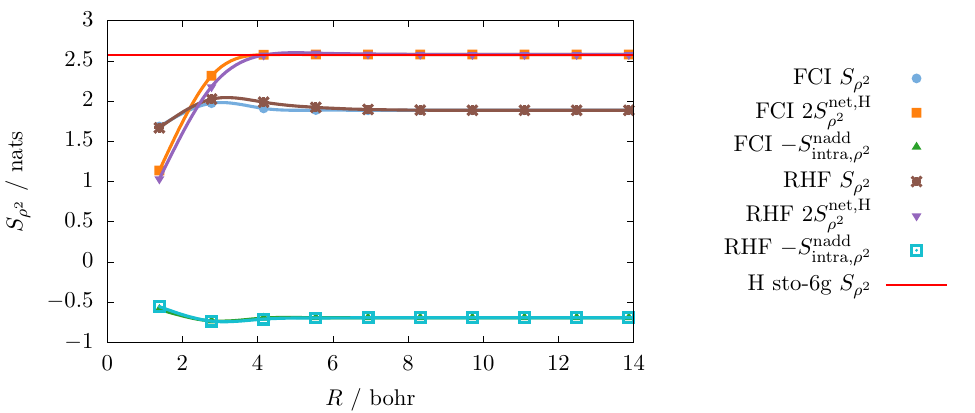}
    \caption{Net, nonadditive-intra ($\sum_A p_A \log p_A$) and total Rényi $\alpha = 2$ entropy terms for H\textsubscript{2} at minimal-basis (STO-6G) for different methods at various internuclear distances (points) and entropy of the isolated atoms (lines). The lowest $R$ value is the equilibrium internuclear distance. Notice how the full entropy does not tend to the net entropy, but to the net entropy plus the nonadditive-intra entropy, as predicted algebraically. It also does not coincide with the H isolated atom Rényi entropy, again due to the presence of the nonadditive-intra contribution of $-\log(2)$. The total net entropy of the molecule equals the entropy of a single atom, due to the weighting of the net entropies using the $p_A$ factors. 
    We use $S_{\rho^2} \equiv S_{\rho}^{\alpha=2}$, and likewise for the terms arising from the entropy decomposition.}
    \label{fig:h2-renyi-minimal}
\end{figure}

\begin{figure}[h!]
    \centering
    \includegraphics[width=0.8\textwidth]{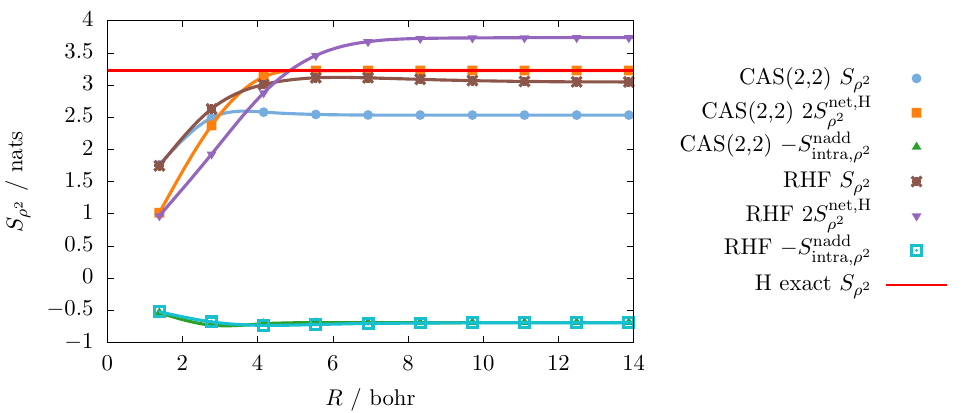}
    \caption{Net, nonadditive-intra ($\sum_A p_A \log p_A$) and total Rényi $\alpha = 2$ entropy terms for H\textsubscript{2} at aug-cc-pVTZ basis for different methods at various internuclear distances (points) and entropy of the isolated atoms (lines). The lowest$R$ value is the equilibrium internuclear distance. Notice how the full entropy does not tend to the net entropy, but to the net entropy plus the nonadditive-intra entropy, as predicted algebraically. It also does not coincide with the H isolated atom Rényi entropy, again due to the presence of the nonadditive-intra contribution of $-\log(2)$. The total net entropy of the molecule equals the entropy of a single atom for the correlated CAS(2,2) calculation, but not in the HF case.
    We use $S_{\rho^2} \equiv S_{\rho}^{\alpha=2}$, and likewise for the terms arising from the entropy decomposition.}
    \label{fig:h2-renyi-aug}
\end{figure}

Numerical results for Rényi $\alpha = 2$ entropies are included in Figure \ref{fig:h2-renyi-minimal} for minimal-basis H\textsubscript{2} and in Figure~\ref{fig:h2-renyi-aug} for augmented basis H\textsubscript{2}. 
The numerical results support our findings. 
Notice also how the convergence to the infinite distance limit is much faster now than in the Shannon entropy case, reaching the limiting value at smaller distances. Also, both bases give much closer entropy terms (and total entropies) than their Shannon counterpart. Thus, we expect, for high values of $\alpha$ and when minimal basis sets are used, that the entropies reduce to their infinite distance counterparts in a way that is much less sensitive to the interatomic distance and the theoretical method used to obtain the densities. 
Here too, the use of augmented basis has a variational effect that in some way discerns the theoretical method, giving a correct atomic limit only for the  CAS(2,2) level, but not for the Hartree-Fock calculation. This is due to the predicted variational deformation of the large distance limit atoms in the Hartree-Fock case.

\section{Methods}
Geometry optimizations and electronic energy calculations were carried out with the Gaussian16 Rev.A.03 \cite{gaussian} quantum chemistry package, while valence-bond (Heitler–London) calculations were performed using GAMESS (US) 2020 R2 \cite{GAMESS}. Numerical integrations were conducted with our in-house code PROMOLDEN.

Following Becke's multicenter integration scheme \cite{becke}, the integral of an arbitrary function $F({\bf r})$ over all space was partitioned into atomic contributions:
\begin{equation}
\label{eq:integration}
\int F({\bf r}) \, d{\bf r} = \sum_A \int \omega_A({\bf r}) F({\bf r}) \, d{\bf r} = \sum_A \langle F \rangle_A,
\end{equation}
where the sum extends over all atoms $A$ in the molecule, and the Becke weight functions $\omega_A({\bf r})$ satisfy $\sum_A \omega_A({\bf r}) = 1$ at every point in space. These weights were constructed using Becke's original recipe.

Each atomic contribution $\langle F \rangle_A$ was evaluated numerically using a product grid comprising 1000 radial points and 434 angular points. The radial coordinate was transformed to an auxiliary variable $q \in [-1, +1]$ via
${\bf r} = R \left( \frac{1+q}{1-q} \right)$,
where $R$ is the Bragg radius of the atom and the $q$ points were chosen as $q_i = \cos\left( \frac{i\pi}{n+1} \right)$ with $n = 1000$. Angular integration employed the popular Lebedev-Laikov quadrature \cite{lebedev}.


\section{Conclusion}

In this work, we have systematically examined the behavior of entropic measures derived from the electron density and related affine objects, combining algebraic analysis with numerical integration. Particular emphasis was placed on static correlation and extensivity, assessed through the infinite internuclear-distance limits of these measures. Our study focused on the Shannon entropy and the $\alpha$-order Rényi entropies, investigating their decomposition under a Mulliken-type atomic partition into additive (net atomic and overlap) and nonadditive contributions.

For minimal-basis calculations, we showed that the Shannon entropy does not encode the degree of electronic correlation present in the underlying density. In this case, the expected atomic limit—where the molecular entropy at large bond distances approaches the sum of isolated atomic entropies—is satisfied irrespectively. In contrast, the corresponding shape-function entropy violates this extensivity even in a minimal basis. Rényi entropies with $\alpha \neq 1$ also fail to reach the atomic limit, owing to a persistent nonadditive contribution that depends explicitly on $\alpha$.

Extending these analyses to a larger basis set (aug-cc-pVTZ) revealed that only sufficiently correlated densities, such as those from full-valence CAS calculations, recover the appropriate large-distance limits (atomic or atomic + nonadditive). Uncorrelated Hartree–Fock densities consistently overestimate the entropy, as confirmed for H\textsubscript{2}, N\textsubscript{2}, and H\textsubscript{2}O. This behavior stems from increased variational flexibility rather than from improved accuracy of the electronic energy.

By establishing a rigorous atom-based entropic partition and deriving general algebraic asymptotic limits for both density- and shape-function–based entropic forms, we demonstrate that extensivity poses a fundamental challenge for these measures— particularly for shape-function entropies—and that static correlation is reflected only in an incomplete form. These findings support the adoption of more sophisticated entropic descriptors, constructed from higher-dimensional Hilbert-space objects, when investigating electronic-correlation effects in molecular systems.

\section*{Acknowledgements}
The authors acknowledge the Spanish MICINN (Grant No. PID2024-155569NB-I00, https:// doi. org/ 10. 13039/ 50110 0011033) and the European Union “ERDF A way of making Europe” for financial support. D.J.L.R. is financed by Fundação para a Ciência e Tecnologia (FCT), through PhD grant 2023.00844.BD.

\section*{Conflict of interest}
The authors declare no conflict of interest. 

\selectlanguage{english}

\bibliographystyle{naturemag} 
\bibliography{article.bib}

\end{document}